# Superconducting Nanowire Single-Photon Detectors for Quantum Information

Lixing You

[1]State Key Laboratory of Functional Materials for Informatics, Shanghai Institute of Microsystem and Information Technology (SIMIT), Chinese Academy of Sciences, Shanghai 200050, China

[2]Center of Materials Science and Optoelectronics Engineering, University of Chinese Academy of Sciences, Beijing 100049, China

[3]CAS Center for Excellence in Superconducting Electronics (CENSE), Shanghai 200050, China

[4] Zhejiang Photon Technology Co Ltd, Guigu Science Park, Jiashan, Zhejiang 31400, China

*Corresponding author: *lxyou@mail.sim.ac.cn

**Abstract:**

The superconducting nanowire single-photon detector (SNSPD) is a quantum-limit superconducting optical detector based on the Cooper-pair breaking effect by a single photon, which exhibits a higher detection efficiency, lower dark count rate, higher counting rate, and lower timing jitter when compared with those exhibited by its counterparts. SNSPDs have been extensively applied in quantum information processing, including quantum key distribution and optical quantum computation. In this review, we present the requirements of single-photon detectors from quantum information, as well as the principle, key metrics, latest performance issues and other issues associated with SNSPD. The representative applications of SNSPDs with respect to quantum information will also be covered.

## 1. Introduction:

Superconductivity, which was discovered by the Dutch physicist Heike Kamerlingh Onnes on April 8, 1911, is one of the most renowned macroscopic quantum effects [1]. Various applications of superconductors have been applied and demonstrated in the past century with the increasing understanding of superconductivity. For example, superconducting magnets have been widely applied in commercial magnetic resonance imaging machines and several major science projects, including the International Thermonuclear Experimental Reactor [2] and superconducting maglev [3]. A superconducting quantum interference device is one of the most sensitive magnetic devices for biomagnetism and geophysics exploration [4; 5]. Superconductors can be used to achieve sensing and detection because of their many extraordinary properties, including zero resistance, the Josephson effect, and the Cooper-pair.

A photon is the quantum of an electromagnetic field, including electromagnetic radiation, such as light and radio waves, which is an elementary particle with a certain energy $E = h\upsilon = hc/\lambda$, where $h = 6.626 \times 10^{-34}$ $J{\cdot}s$ is the Planck's constant, $c = 2.998 \times 10^8$ $m/s$ is the speed of light in vacuum, $\upsilon$ and $\lambda$ are the frequency and wavelength of the photons, respectively. Further, microwave photons, terahertz photons, visible/near-infrared (NIR) photons, and high-energy photons/particles exist. Superconductors can be engineered for detecting photons. Various superconducting sensors and detectors have demonstrated an unparalleled performance for almost the whole electromagnetic spectrum from a low-frequency microwave to a high-energy particle. In this review, we focus on the detection of traditional photons for NIR and visible wavelengths (400–2000 nm), having a typical energy of approximately $1–5 \times 10^{-19}$ J (0.6–3 eV). NIR and visible lights are popular bit carriers for optical communication. NIR and visible photons are also among the key quantum bit carriers of quantum information (QI).

The photon detection mechanism of superconductors can be simple and straightforward. According to the BCS theory developed by Bardeen, Cooper, and Schrieffer in 1957 [6], pairs of electrons (Cooper-pairs) are formed in superconductors via electron–phonon interaction when the temperature $T$ is lower than its superconducting transition temperature $T_c$. A Cooper-pair has the minimum binding energy $E_g = 2\Delta(T)$, where $\Delta(T)$ is the energy gap of the superconducting material, which is sensitive to $T$. When $T << T_c$, $E_g = 2\Delta(0) = 3.528 k_B T_c$, where $k_B = 1.381 \times 10^{-23}$ J/K is the Boltzmann constant. If a photon is absorbed by a superconductor, then it may break a Cooper-pair and produce two quasi-particles if the photon energy $E(\lambda)$ is larger than $E_g$. Let us consider a traditional low-temperature niobium nitride (NbN) superconductor for calculation purposes. NbN has a $T_c$ of 16 K, corresponding to the $\Delta(0)$ of 3.2 meV. Theoretically, a single NIR photon having a wavelength of 1550 nm and an energy of 0.8 eV can break 125 Cooper-pairs. If the Cooper-pair breaking event can produce a measurable physical quantity, which can be captured using an appropriate instrument, then a single-photon detection event is registered.

Different types of superconducting single-photon detectors (SPDs) exist. They may have different operation principles, use different device structures and materials, and generate different output signals even though they rely on the Cooper-pair breaking mechanism. According to different operation principles, the superconducting SPDs can be divided into different types [7]: transition edge sensor (TES), superconducting tunnel junction (STJ), microwave kinetic inductance detector (MKID), and superconducting nanowire single-photon detector (SNSPD).

The TES usually comprises an ultralow-temperature superconducting film, such as tungsten (W), which produces a measurable resistive change within a sharp normal-to-superconducting transition upon photon absorption. The TES exhibits high detection efficiency, low speed, high timing jitter, and unique photon number resolvability and usually requires a sub-Kelvin operating temperature [8; 9]. When an STJ operates as an SPD, one superconducting film (electrode) absorbs the photons and the photon energy is converted into broken Cooper-pairs and phonons. The transfer of the charge carriers from one electrode to another will result in a measurable electrical current on the STJ [10]. An MKID is a thin-film high-Q superconducting micro-resonator, the resonance frequency and internal quality factor of which change when the incoming photons break the Cooper-pairs in the superconductor. The frequency shift and internal dissipation signal measurements are referred to as the frequency readout and dissipation readout, respectively [11; 12]. STJ and MKID exhibit an optical photon detection ability; however, no practical detectors have yet been developed. Meanwhile, an SNSPD usually has a nanowire/nanostrip structure. When a photon is absorbed by the current-biased SNSPD, a local resistive domain can be observed, resulting in a voltage pulse, which indicates a detection event. An SNSPD has high detection efficiency, low dark count rate, high speed, and low timing jitter [13].

Semiconducting SPDs, such as the single-photon avalanche diode (SPAD), have been widely applied. SPADs are avalanche photodiodes biased at fields above an avalanche breakdown in Geiger mode, where a self-sustaining avalanche current can be triggered by an incident single photon.

Compacted Si SPADs having a detection efficiency of more than 70% are commercially available for the detection of visible photons. InGaAs/InP SPADs are produced for detecting NIR photons because InGaAs has a lower bandgap than Si. However, their detection efficiency is usually not more than 30%, and the dark count rate is several tens of thousands of counts per second. A tricky NIR photon detection module can be referred to as the upconversion SPD. The upconversion SPD utilizes sum frequency generation in a periodically poled lithium niobite waveguide or bulk crystals, converting the NIR photons into shorter-wavelength photons and detecting them using a Si SPAD, which may increase the detection efficiency to become more than 40%.

Quantum mechanism and information science are two significant scientific revolutions in the 20th century [14]. Although many innovations based on quantum mechanism have been successfully applied in information science and technology (e.g., lasers and transistors), one does not see direct control or manipulation of the quantum states at the quantum level. In the previous few decades, modern science and technology has enabled the control and manipulation of various quantum systems along with information science, producing an emerging field: **QI**. The science and technology of QI can produce revolutionary advances in the fields of science and engineering, involving communication, computation, precision measurement, and fundamental quantum science. This is usually called "the second quantum revolution." The considerable potential of QI is expected to attract research funds of tens of billions of US dollars over the next few years from governments of many countries and regions, including Australia [15], Canada [16], China [17], Europe [18], Japan [19], Russia [20], the United Kingdom [21], and the United States [22]. Many renowned information technology companies, such as Google, IBM, Microsoft, Huawei and Alibaba are also participating in the QI race.

As an emerging and fast-growing field, QI does not have a uniform definition and classification. Different countries also have different classifications with respect to their initiatives. From the application perspective, the QI technology has three directions, including quantum communication, quantum computation/simulation, and quantum measurement/metrology. Quantum sensors and detectors are the core devices of the QI systems. An SPD is essential for photon (optical quantum) measurement/metrology or any measurement/metrology, where the signal can be converted to photons. For achieving quantum communication and computation, an SPD will play an irreplaceable role in the QI systems as long as a photon functions as the quantum state carrier [23].

Table 1 summarizes the state-of-the-art performances of various SPDs for QI at a telecom band wavelength of 1550 nm, which is one of the key wavelengths for quantum communication and optical quantum computation. Different SPDs have been reviewed by Hadfield in 2009 [24] and Eisaman et al. in 2011 [25]. The data in Table 1 indicate that SNSPDs surpass their counterparts as an excellent SPD candidate for QI experiments. Many applications have been demonstrated using SNSPDs. Commercial SNSPDs are produced by several companies. Encouraged by the major success of the SNSPDs in QI, this review focuses on the SNSPDs and their applications in QI, although other impressive applications, such as deep-space communication [26] and light detection and ranging [27; 28], are available. Furthermore, some general or specific reviews of SNSPDs can be found [29-33].

This review introduces the SPD requirements based on QI, the state of the art in SNSPDs, and the applications of QI, which will provide readers with a broad perspective from the application perspective. This study contains six chapters. Chapter 2 focuses on the SPD requirement based on QI. Chapter 3 introduces SNSPDs and the operation principle, performance parameters, key issues, and latest achievements related to SNSPDs. Chapter 4 summarizes the SNSPD applications with respect to QI. Chapter 5 presents other issues, including standardization and business market. For clarity and completeness, we will also refer the interested readers to a more specialized literature on different topics.

Table 1. State-of-the-art performance of different SPDs at a wavelength of 1550 nm for quantum information

| Detector type | Detection efficiency | Dark count rate | Counting rate | Timing jitter | Operation temperature | Note |
|---|---|---|---|---|---|---|
| InGaAs/InP SPAD | 27.5% | 1.7 kcps | >10 MHz | / | 223 K | [34], with an after-pulse probability of 9.1% |
| Upconversion SPD | 59% | 460 kcps | 5 MHz | / | 300 K | [35] |
| Upconversion SPD | 28.6% | 100 cps | / | / | 300 K | [36] |
| W TES | 95% | / | ~1 MHz | / | 0.1 K | [8] |
| NbN SNSPD | 90% | 10 cps | ~30 MHz | 70 ps | 2.1 K | [37; 38] DE is 95% (unpublished data from author) |
| WSi SNSPD | 93% | 1000 cps | ~10 MHz | 150 ps | 0.12 K | [39] |
| MoSi SNSPD | 87%/98% | 100 cps | ~10 MHz | 76 ps | 0.7 K | [40]; 98% efficiency was reported [41] without other parameters |

## 2. SPD requirements from QI

### 2.1 Quantum communication

Cryptography is the core of security communication. Information-theoretical secure communication can be achieved using the one-time-pad method, and the key must be as long as the message and cannot be reused [42; 43]. The manner in which such a long key can be distributed in the presence of an eavesdropper is called the key distribution problem. Quantum key distribution (QKD), the core of quantum communication, is developed to solve this central challenge. QKD provides information-theoretical security based on the basic principle of quantum physics. For comparison, the traditional cryptography method is based on computation complexity, and its security is dependent on the algorithm or computational power. QKD or quantum communication is believed to be the first commercial application of quantum physics at the single quantum level. The first QKD protocol was proposed by Bennett and Brassard in 1984 and is referred to as BB84 [44]. Subsequently, many other protocols were proposed and demonstrated to make QKD more practical and powerful against various possible attacks, including decoy-state QKD [45-47], measurement device-independent QKD (MDI–QKD) [48], and the latest two-field QKD (TF–QKD) [49]. For obtaining the details of quantum cryptography, refer to the review article by Gisin et al. in 2002 [50] and the latest review by Xu et al. in 2019 [51].

We may consider a classical QKD system using the BB84 protocol as an example. A sequence of single photons carrying qubit states is sent to Bob by Alice through a quantum channel, shown in Figure 1(a). Alice adopts the polarization states of single photons to encode random bits. Bob randomly selects the measurement bases, either rectilinear or diagonal, to perform measurements using two SPDs. The sifted key is obtained by keeping only the polarization data encoded and detected in the same basis. Alice and Bob can share the final secret key after performing additional classical post-processing on the sifted key. A QKD system (Figure 1(b)) needs some key components, including single-photon sources, quantum channel, and SPDs. Two or four SPDs are necessary to implement the BB84 QKD system. Regardless of the manner in which the QKD protocols are developed, SPDs play an irreplaceable part in the QKD systems if a single photon is used to transmit the quantum state. The performance matrix of the QKD systems, such as the maximum transmission distance, secure key rate ($R_{SK}$), and quantum bit error rate ($R_{BE}$), is dependent on the performance of the SPDs. Simple equations for the decoy-state BB84 QKD are as follows:

$$\begin{cases} R_{SK} \propto \eta \cdot f \cdot u \cdot L \\ R_{BE} \propto R_{dc}/R_{SK} \end{cases}, \tag{1}$$

where $\eta$ is the detection efficiency of the SPD, $f$ is the clock frequency, $u$ is the average photon number per pulse, $L$ is the total channel loss, and $R_{dc}$ is the dark count rate of SPD. Equation 1 indicates that the performance of the QKD system is dependent on the key parameters of the SPDs, including $\eta$ and $R_{dc}$. The dead time and timing jitter (TJ) of SPDs will affect the performance of the high-speed QKD system.

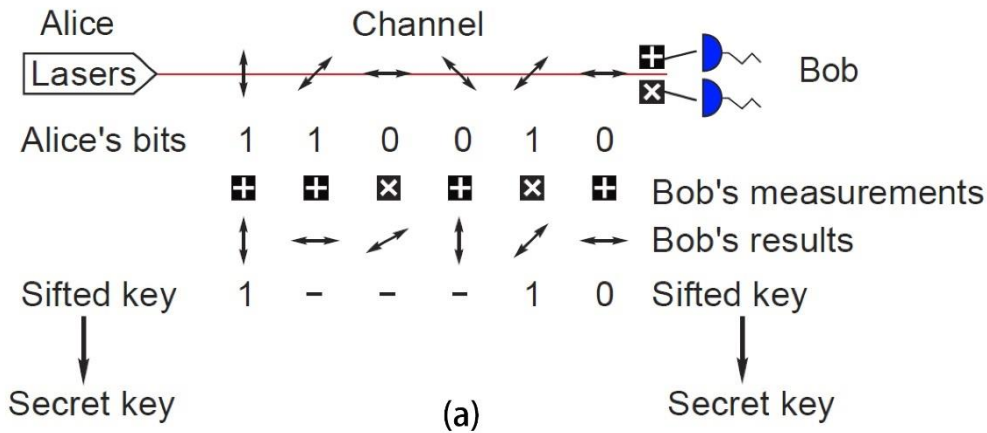

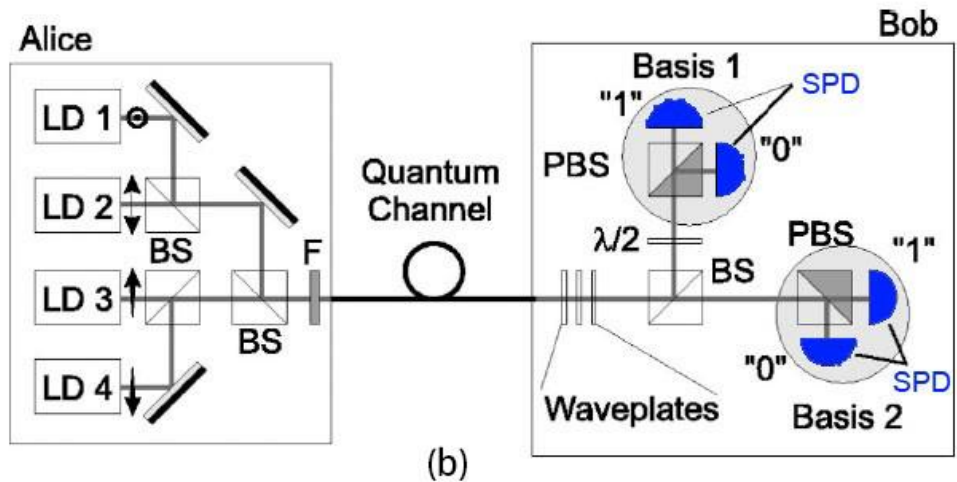

Figure 1. (a) Schematic of the BB84 protocol. Two or four SPDs are required in the system with/without a polarization modulator. Reproduced with permission. [51] Copyright 2019, the authors (arXiv); (b) A typical QKD system with a decoy-state BB84 protocol using polarization coding. LD, laser diode; BS, beam splitter; F, neutral density filter; PBS, polarizing beam splitter; l/2, half waveplate. Reproduced with permission. [50] Copyright 2002, American Physical Society.

**2.2 Quantum computation**

Quantum computers use quantum superposition to process information in parallel. Thus, they have a fundamental computing advantage over classical computers and can decrypt most of the modern encrypted communication. Some prototype quantum computation systems have been developed based on the theoretical and experimental studies conducted over the previous two decades. At the end of 2019, Google initially demonstrated quantum advantage/supremacy using a processor containing programmable superconducting quantum bits (qubits) to create quantum states on 53 qubits [52]. Noisy intermediate-scale qubits will be useful for exploring many-body quantum physics. They may also have other useful applications. However, the 100-qubit quantum computer will not change the world immediately; we should consider it as a significant step toward achieving more powerful quantum technologies in the future [53]. Unlike the quantum communication mostly based on photon/optical quantum, quantum computers can be built using several different physical qubits, including superconducting circuits [52], trapped ions [54], quantum dots [55; 56], nuclear magnetic resonance [57], nitrogen-vacancy centers in diamond [58], and photons [59]. In addition to the superconducting quantum computer, the optical quantum computer has approached a milestone in terms of quantum advantage. Lu et al demonstrated boson sampling with 20 input photons and a 60-mode interferometer in a $10^{14}$-dimensional Hilbert space, which is equivalent to 48 qubits [60].

To build an optical quantum computer, one needs indistinguishable single photons, low-loss photonic circuits, and high-efficiency SPDs. Figure 2 shows an experimental setup of boson sampling using the optical quantum computer [60]. A single InAs/GaAs quantum dot, resonantly coupled to a microcavity, is used to create pulsed resonance fluorescence single photons. For demultiplexing, 19 pairs of Pockels cells and polarized beam splitters are used to actively translate a stream of photon pulses into 20 spatial modes. Optical fibers with different lengths are used to compensate time delays. The 20 input single photons were injected into a 3D integrated, 60-mode ultra-low-loss photonic circuit comprising 396 beam splitters and 108 mirrors. Finally, the output single photons were detected by 60 SNSPDs. All the coincidences were recorded using a 64-channel coincidence count unit (not shown in Figure 2).

The coincidence counting (CC) of the *n*-photon boson sampling can be expressed as

$$CC(n) = \frac{R_{pump}}{n} \cdot \left(\eta_{QD} \cdot \eta_{de} \cdot \eta_C \cdot \eta_{SPD}\right)^n \cdot S, \qquad (2)$$

where $R_{pump}$ is the pumping repetition rate of the single-photon source, $\eta_{QD}$ is the single-photon source brightness, $\eta_{de}$ is the demultiplexing efficiency of each channel, $\eta_c$ is the average efficiency of the photonic circuit, $\eta_{SPD}$ is the detection efficiency of the SPDs, and *S* is the ratio of no-collision events to all possible output combinations. Equation 2 shows that a high $\eta_{SPD}$ of approximately 1 is critical for boson sampling when *n* is large. Increasing $\eta_{SPD}$ from 0.3 (typical value for SPAD) to 0.8 (typical value for SNSPD) at a wavelength of 1550 nm can improve *CC (n = 50)* by 21 orders of magnitude. Thus, the sampling time can be considerably reduced. Furthermore, SPDs should be sufficiently fast to match the $R_{pump}$ of the single-photon source; otherwise, $\eta_{SPD}$ cannot be guaranteed.

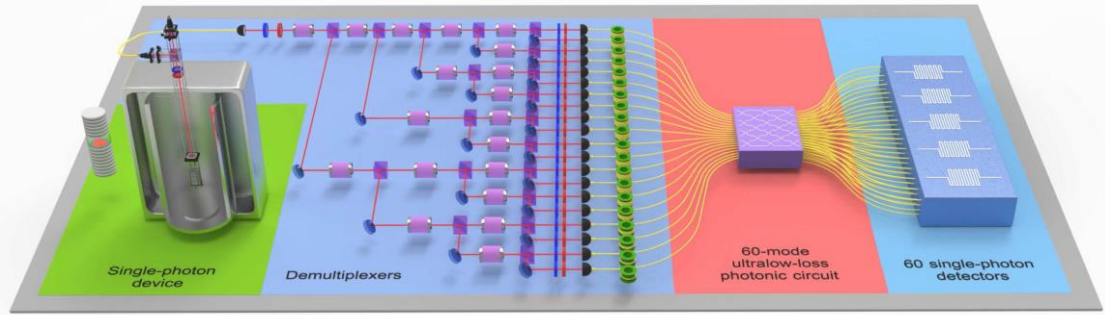

Figure 2. Experimental setup of boson sampling using 20 photons. The setup includes four key parts, i.e., a single-photon device, demultiplexers, a photonic circuit, and SNSPDs. Reproduced with permission. [60] Copyright 2019, American Physical Society.

## 2.3 SPDs for QI

The previously conducted analysis indicates that SPDs with high detection efficiency, low dark count rate, high counting rate, and low TJ are indispensable in case of QI. Semiconducting SPADs were previously widely applied to QI. However, their performance cannot keep up with the pace of QI development. SNSPDs were initially demonstrated in 2001 [13]. Their performance has considerably improved over the previous two decades, considerably advancing QI science and technology as a key enabling technology.

## 3. SNSPD

### 3.1 History

An SNSPD based on the non-equilibrium hotspot effect observed in ultrathin superconducting films was first proposed by Kadin et al. in 1996 [61]. The concept of SNSPD was successfully demonstrated by Gol'tsman using the NbN strip (200-nm wide and 5-nm thick) [13]. Although the first result of SNSPD was moderately satisfactory, it attracted the attention of the superconducting electronics and QI communities. The first experimental demonstration of SNSPD with respect to QKD was conducted by Hadfield et al. in 2006 [62]. The considerable potential of SNSPDs for various applications, including QI, has fascinated researchers all over the world. The performance of SNSPDs has been effectively enhanced based on improved knowledge and experience in materials, design, fabrication process, and the theory. Numerous milestones in QI experiments have been achieved using SNSPDs with unparalleled parameters, such as loophole-free tests of local realism [63], QKD with over 500 km of optical fiber [64; 65], and boson sampling with 20 input photons [60]. Another successful milestone that was achieved outside the scope of QI is deep-space communication [26]. In 2013, NASA achieved laser communications between a lunar-orbiting satellite and ground stations on Earth with downlink data rates of up to 622 Mb/s utilizing SNSPDs at a wavelength of 1550 nm. An increasing number of exciting achievements in QI will be attained in the foreseeable future.

### 3.2 Detection mechanism

The microscopic detection mechanism of SNSPD is complicated and not well understood. Various relevant studies have been conducted using different theoretical models and from different aspects. No unified model could explain all the experimental results, and some experimental results are not consistent. Regardless, we briefly explain the detection process using Cooper-pair breaking and an electrothermal feedback model [66] without considering the detailed microscopic mechanisms [67; 68].

Let us begin with a typical SNSPD, which is usually a nanowire/nanostrip with a width of approximately 100 nm made of ultrathin (5–10-nm thick) superconducting film (such as NbN and WSi). An SNSPD is usually cooled to a temperature of less than $0.5T_c$ and current-biased with a bias value close to but smaller than its switching current $I_{sw}$ such as $0.9I_{sw}$ (Figure 3[a]-i). The switching current $I_{sw}$ is defined as the maximum current with which the nanowire can sustain superconductivity. When a photon is absorbed by the nanowire, it may break hundreds of Cooper-pairs because a single photon's energy (~1 eV) is usually two to three orders magnitude higher than the binding energy of a Cooper-pair (say, 3.2 meV for NbN). Thus, the depaired quasi-particles form a hotspot in the nanowire, which repels the supercurrent into the superconducting path around the hotspot (Figure 3[a]-ii). Then, the current around the hotspot may exceed the switching current, enabling the hotspot to expand and form a resistive slot across the nanowire (Figure 3[a]-iii, iv). The resistive slot (usually several hundred ohms) will grow along the direction of the nanowire (Figure 3[a]v) due to the Joule heating effect. Simultaneously, the current in the nanowire will be forced to flow into the readout circuit; thus, a measurable signal can be obtained. With the reducing bias current in the nanowire and thermal relaxation via the substrate, the resistive slot will cool and finally disappear (Figure 3[a]-vi); then, the SNSPD is ready for the next incoming photon. Figure 3(b) gives a simple circuit by assuming that SNSPD is a switch in parallel with a resistor ($R_n(t)$) and in series with an inductor. The switch is used to simulate the detection event triggered by a photon. $R_n(t)$ simulates the dynamic resistive slot with a time-dependent resistance. Generally, the total inductance of an SNSPD comprises magnetic (geometrical) inductance and kinetic inductance ($L_k$). The kinetic inductance is described by the imaginary part of the complex conductivity in the superconducting state, which is considerably larger than its geometric inductance. Therefore, $L_k$ is often adopted to represent the total inductance of SNSPDs [69]. This circuit will give a voltage pulse shown in Figure 3(c), which usually has a sharp rising edge with a time constant $\tau_1 = L_k/(Z_0+R_n)$ and a slow falling edge with a time constant $\tau_2 = L_k/Z_0$, where $Z_0 = 50\ \Omega$ is the impedance of the readout circuit. $\tau_1$ is usually approximately 1 ns and $\tau_2$ is in the range of some nanoseconds to a few hundred nanoseconds, which is related to the active area size of the SNSPD.

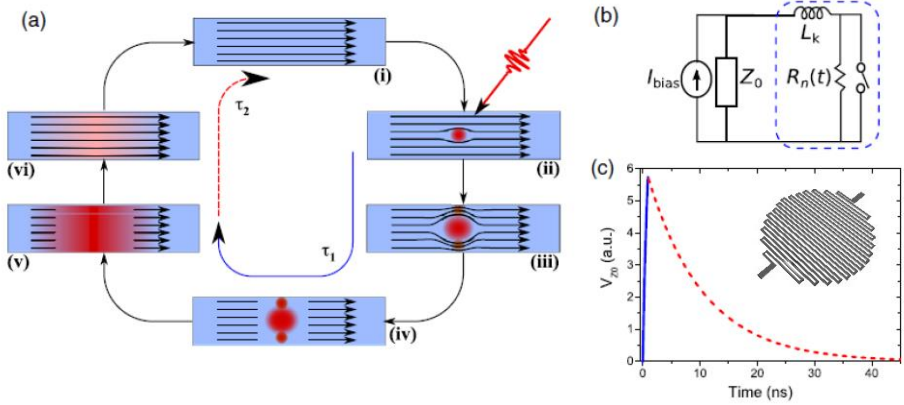

Figure 3. (a) Schematic of the detection cycle; (b) a simplified circuit model of an SNSPD; and (c) the output signal of an SNSPD upon a detection event [31]. The inset added to the original reference is a schematic of a meandered SNSPD. Reproduced with permission [31]. Copyright 2012, IOP Publishing Ltd.

Although the above schematics are simple, they can generally explain how the SNSPD works and what happens after the detection of a photon, which can facilitate understanding from the application perspective. However, many parameters must be designed and tuned carefully to make a functional SNSPD with respect to different aspects, such as materials, geometrics, circuits, and operation parameters.

Figure 4(a) shows the scanning electron microscopy (SEM) image of an SNSPD with an optical cavity on top. The NbN film was structured into meandered nanowires using electron-beam lithography and reactive ion etching. The upper-left image shows the edge of the active area. The meandered nanowire has a width of 100 nm and a filling ratio of 0.5. It usually has a round corner that turns back and forth over the active area. The round corner structure avoids the current crowding effect, which may reduce the switching current of the SNSPD [70; 71]. The upper-right image is an optical image of a packaged SNSPD, with a fiber aligned vertically from the top [72]. Figure 4(b) shows the raw amplified signal of single photon detection recorded by an oscilloscope, which is consistent with the simulation result shown in Figure 3(c).

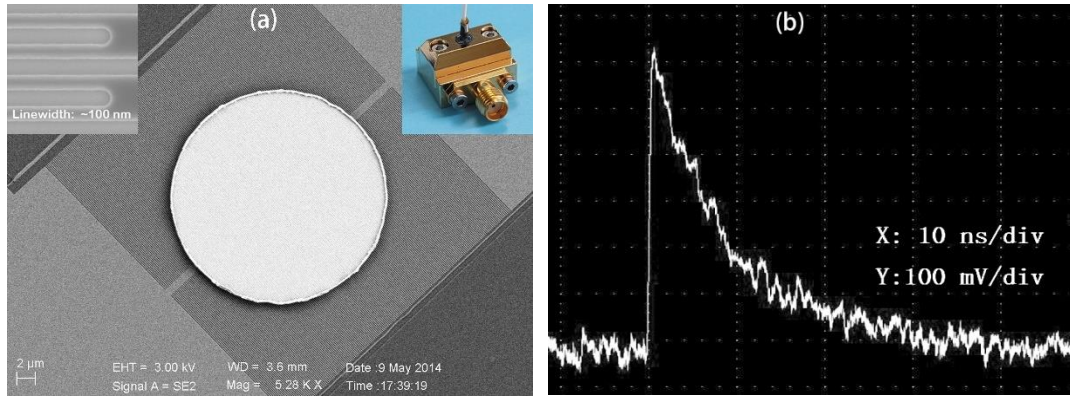

Figure 4. (a) SEM image of a meandered SNSPD with an optical cavity on the top. The left-top inset that shows the nanowire has a line width of 100 nm, and the right inset shows an optical image of a packaged SNSPD with fiber-alignment; (b) oscilloscope single shot of a single-photon response signal.

3.3 Metrics of SNSPD

- Detection efficiency

The system detection efficiency (SDE) is a term that is easily accepted by users. SDE is defined as a parameter that indicates how effectively the SNSPD system can detect photons, including all the losses inside the system. SDE can be usually presented as $SDE = \eta_{coupling} \cdot \eta_{absorption} \cdot \eta_{intrinsic}$, where $\eta_{coupling}$ is the efficiency of the incident photons being coupled to the active area of a detector, $\eta_{absorption}$ is the absorption efficiency of the photons coupled to a detector, and $\eta_{intrinsic}$ is the triggering efficiency of the absorbed photons that can produce a detectable electric signal. The ideal SNSPD system should have an SDE of unity. It is difficult to practically achieve this condition. If all the three $\eta$s values are 0.97, the final SDE is 0.91. In addition, an excellent commercial fiber connector with a loss of 0.1 dB inside the system can result in an efficiency loss of 0.02.

Three different optical structures have been developed for SNSPDs. The most popular one is the vertical-coupling method, which also provides the highest SDE as a standalone detector (Figure 4). The second is the waveguide-coupled method, with superconducting nanowires fabricated on the top of the waveguide. Waveguide-coupled SNSPDs generated modest SDEs due to the coupling loss of the waveguide; however, $\eta_{absorption}$ and $\eta_{intrinsic}$ were claimed to be close to 1, which may play an important role in integrated quantum photonics [32; 73]. The final one is the microfiber coupling method, which utilizes the evanescent field of microfiber when the superconducting nanowires are in close contact with the microfiber. Microfiber coupling SNSPDs obtained an SDE of more than 50% for the broadband spectrum from visible to NIR, possibly finding interesting applications in sensing and spectrometry [74; 75]. This review will focus on the vertically coupled SNSPDs.

To achieve a high coupling efficiency, we should make the active area large enough to enable the effective coupling of the incoming photons. Various optic methods, such as lens, are helpful for enhancing the coupling. Several packaging methods have been developed for effective coupling, including using the cryogenic nanopositioner and novel self-alignment technique [30], with which the loss can become as low as 1% [76].

High absorption is more challenging than high coupling for SNSPDs. Previous SNSPDs made of ultrathin films with a simple meandered nanowire structure exhibit a low absorption of 0.3 (OS1 in Figure 5(a)). Rosfjord et al. initially introduced a cavity structure to effectively enhance the absorptance (OS2 in Figure 5(a) [77]. To further improve the absorptance, more sophisticated cavities were designed, such as a double cavity structure with backside illumination (OS3 in Figure 5(a) [78] and a dielectric mirror (distributed Bragg reflector) with frontside illumination (OS4 Figure 5(a) [38]. In principle, the optimized OS3 and OS4 may achieve an absorption of more than 0.99 according to simulation. However, based on the simulation results, imperfect materials, the deviation of the geometric parameters of the optical materials, and superconducting nanowires may reduce the absorption. A multilayer design may help to reduce the influence of the aforementioned imperfections [79]. The interface reflection at the surfaces of the fiber tip and detector chip slightly reduces the absorption.

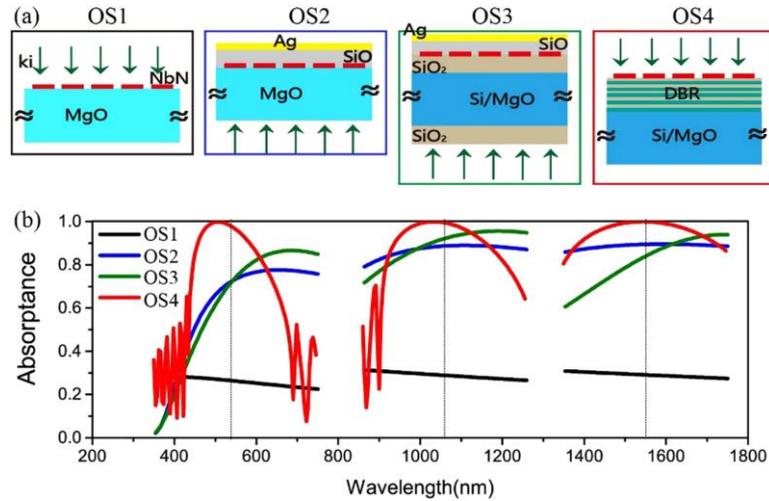

Figure 5. (a) OS1–OS4: Four kinds of optical structures used for vertically coupled SNSPD. (b) The calculated optical absorptance for different optical structures at 532, 1064, and 1550 nm. At 532 and 1064 nm, the MgO substrate was adopted in the simulation for OS3 and OS4. At 1550 nm, a Si substrate was used in the simulation of OS3 and OS4. Reproduced with permission [80] Copyright 2017, IOP Publishing Ltd.

$\eta_{intrinsic}$ is directly and closely dependent on the superconducting quality, geometric design, and fabrication precision of the superconducting nanowires. It will be considerably influenced by the operation parameters (i.e., temperature and bias current). Moreover, the geometric and physical uniformity influence $\eta_{intrinsic}$. Any nanowire defects may result in a reduced maximum bias current, preventing $\eta_{intrinsic}$ from reaching unity. Based on the saturated plateau in SDE dependence of the bias current, $\eta_{intrinsic}$ can reach unity in SNSPDs made of various materials, including WSi [39], MoSi [40] and NbN [38].

The aforementioned analysis indicates that SNSPDs with an SDE of close to unity can be obtained. Table 1 shows SDE values of more than 90% at a wavelength of 1550 nm for the WSi, NbN and MoSi SNSPDs. At the Rochester Conference on Coherence and Quantum Optics in 2019, Reddy et al. reported a MoSi SNSPD with an SDE of 98%, which is the highest value observed in case of SNSPDs [41]. Similar results were obtained for other wavelengths [81]. We believe that a maximum SDE of close to unity will be possible for SNSPDs of WSi and NbN upon further optical optimization; this will also be possible for all the other wavelengths from visible to NIR. However, this is an interesting metrology issue related to the accuracy and uncertainty of the measured SDE because no optical power meter at the quantum level is better than SNSPDs [82].

- DCR

The dark count rate (DCR or $R_{dc}$) is defined as the recorded false counts in unit time with respect to the detection events and represents the noise level of an SNSPD. In the measurement method, DCR is the number of counts in unit time recorded with no illumination. DCR consists of background DCR (bDCR) and intrinsic DCR (iDCR).

As a detector with quantum-limit sensitivity, bDCR is difficult to avoid because photons are present everywhere owing to the presence of thermal electromagnetic radiation in a working condition. According to Planck's law of black-body radiation, matter with a certain temperature will radiate photons because of the thermal motion of the particles in matter. In case of fiber-coupled SNSPDs, the thermal radiation of the fiber at room temperature will produce photons that may transmit through the fiber and that may be detected as background dark count. In addition, any stray light penetrating into the fiber may contribute to bDCR. In case of space-coupled SNSPD, bDCR will be considerably large owing to severe environmental radiation. Thermal radiation has a broadband feature and may be filtered partially by filters. However, all the filters need to be operated at low temperatures (40 K or lower) and with an acceptable low loss. Several techniques, such as cooled SMF [83], standalone fiber optic filters or filter bench [84; 85], and dielectric film filters either on the chip [86] or on the tip of a fiber [87], can effectively reduce bDCR with an acceptable sacrifice in terms of SDE. For an SMF-coupled SNSPD, 80% SDE was achieved at a DCR of 0.5 Hz; this is shown in Figure 6 with a bandpass filter (BPF) being used on the fiber end-face [87].

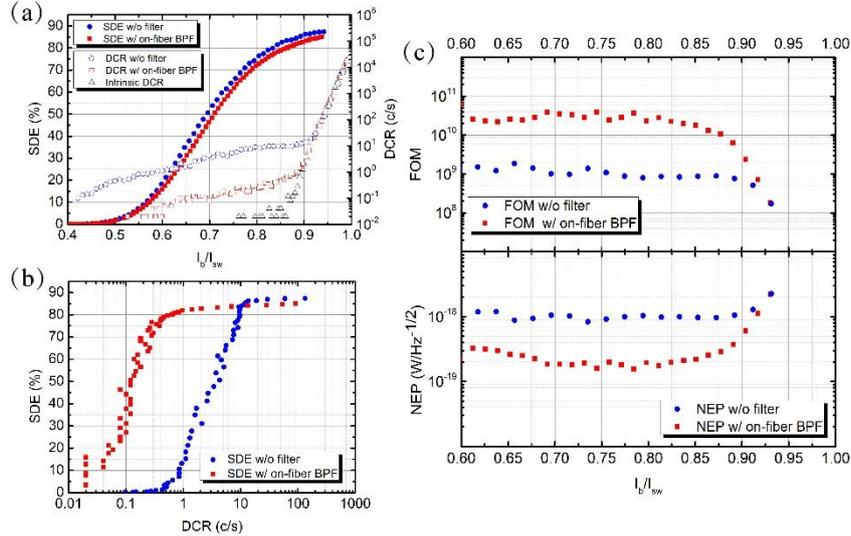

Figure 6. (a) Normalized bias-current dependence of SDE and DCR for an SNSPD coupled using a fiber with and without a BPF on its end-face. The intrinsic DCR of the device is also plotted. (b) SDE as a function of DCR without (blue dots) and with (red squares) a BPF. (c) Top: figure of merit (FOM), FOM = SDE/(DCR × $j_{sys}$). Bottom: noise equivalent power (NEP) as a function of bias current for devices with/without end-face BPF, NEP = $h\nu \cdot \sqrt{2 \cdot DCR}/SDE$ Reproduced with permission [87] Copyright 2018, IOP Publishing Ltd.

When the bias current is high (for example, $I_b/I_{sw}$ > 0.9 in Figure 6[a]), iDCR will be dominant with respect to the DCR. The origin of iDCR is related to the spontaneous vortex motion in the nanowire and is usually exponential to the bias current (black triangles in Figure 6[a]). A few theoretical models (fluctuations of the order parameter, thermally activated and quantum phase slips, and vortex excitations) and experimental studies have investigated the origin of intrinsic dark counts [67]. Although no unified conclusion on the origin has been achieved yet, an SNSPD with a low DCR can be used as long as it is biased not close to $I_{sw}$. iDCR can be neglected when the bias current of an SNSPD is lower than $0.9I_{sw}$, as shown in Figure 6[a].

- Timing jitter

TJ is a key parameter of SNSPDs that surpass that of SPADs and other SPDs. TJ represents the deviation of an ideal periodic single-photon response voltage pulse from the true arrival time. Although it is not yet a critical parameter with respect to the current QI applications, its significance is expected to be revealed soon in high-speed QKD and other applications. To understand the origin of TJ, the contribution of each part to the TJ measurement systems should be understood. The user generally focuses more on the system TJ instead of the detector TJ. The system TJ $j_{sys}$ can be presented as follows:

$$j_{sys} = \sqrt{j_{Int}^2 + j_{SNR}^2 + j_{laser}^2 + j_{SYN}^2 + j_{SPC}^2}, \qquad (3)$$

where $j_{Int}$, $j_{SNR}$, $j_{laser}$, $j_{SYNC}$, and $j_{SPC}$ are the jitters from the SNSPD, the signal-to-noise ratio (SNR) of the output signal, the laser, the synchronization signal of the laser, and the single-photon counting (SPC) module, respectively.

$j_{Int}$ is mainly contributed by the SNSPD geometry. When photons are absorbed at different locations along the long nanowire, the triggered electric signal propagates toward the end of the nanowire with different arrival times. This effect is discovered and verified by a differential cryogenic readout [88]. A long nanowire corresponds to a large geometric contributed jitter. The random time delay between photon absorption and the appearance of the resistive area contributes to the intrinsic jitter [89]. Various types of inhomogeneity, such as gap energy, line width, and thickness, may also produce jitter [90].

$j_{SNR}$ is vital to achieve a practical SNSPD because the amplitude of the original signal at low temperature is ~1 mV. The original signal should be amplified to achieve a high SNR. The slew rate of the rising edge of the signal will also play a role because jitter is measured by calculating the histogram of the timing of a certain voltage on the rising edge. Thus, the device parameters (bias current, kinetic inductance, etc.) and amplifier specifications (gain, noise, and bandwidth) influence jitter [91]. The cryogenic amplifier [81] and impedance-matching taper [92] can also improve SNR, reducing $j_{SNR}$.

$j_{laser}$, $j_{SYNC}$, and $j_{SPC}$ are instrumental specifications. The first two are as small as sub-ps, which are usually neglected. $j_{SPC}$ may vary from a few ps to ~20 ps. The best commercial SPC module may have a full width at half maximum (FWHM) $j_{SPC}$ of <3ps [93]. For research purposes, an oscilloscope embedded with the jitter analysis function (sub-ps jitter) may be adopted to replace the SPC module for reducing the contribution of $j_{SPC}$. However, the data collection time will be much longer than that when the professional SPC module is used [94].

The latest results show that the system TJ $j_{sys}$ can be as low as 3 ps (FWHM). However, it was obtained from a 5-μm-long nanowire, which is not a practical detector because of its low absorption [95]. In case of a practical NbTiN SNSPD having an active area diameter of 14 μm, sub-15 ps $j_{sys}$ was reported with an SDE of ~75% at a wavelength of 1550 nm [81].

● Counting rate

Counting rate (CR) describes how fast the detector can react to the incoming photons, which is the reciprocal of the dead time or the pulse width. The maximum counting rate (MCR) is physically limited to the order of 10 GHz by the thermal relaxation time between the nanowire and the substrate after the absorption of a photon, which is usually a few tens of ps [96; 97]. Practically, SNSPDs have a large kinetic inductance, which slows the process of the bias current's recovery to the initial value after hotspot generation. As estimated from Figure 3(c),

$$\mathrm{CR} = \frac{1}{\tau_1+\tau_2} \approx \frac{Z_0}{L_k}. \tag{4}$$

In case of a NbN SNSPD with a 7-nm thick film, an active area of 15 μm × 15 μm, a fill ratio of 37.5%, and $L_k$ = 1.2 μH, CR is estimated to be ~40 MHz. However, the current recovery to SNSPD is a gradual process, and the SDE also recovers gradually. Thus, the SDE does not completely recover when the next photon arrives at a repetition rate of CR. Hence, the number of counts is usually smaller than the CR·SDE. However, SNSPD may count the incoming photons with a higher repetition rate than CR with a lower efficiency. The useful information for the user is the SDE dependence of the input photons' intensity or the real count rate (Figure 7).

To increase CR, $L_k$ can be reduced by dividing a single nanowire into multiple nanowires. A few different structures, such as arrays [98], interleaved nanowires [99], and parallel nanowires [100; 101], exist. Figure 7 shows a detector with 16 interleaved nanowires that attained a CR of 1.5 GHz with an SDE of ~12% at a photon flux of $1.26 \times 10^{10}$ photons/s. Optimization on circuits may improve the MCR, which is limited by the latching effect [102].

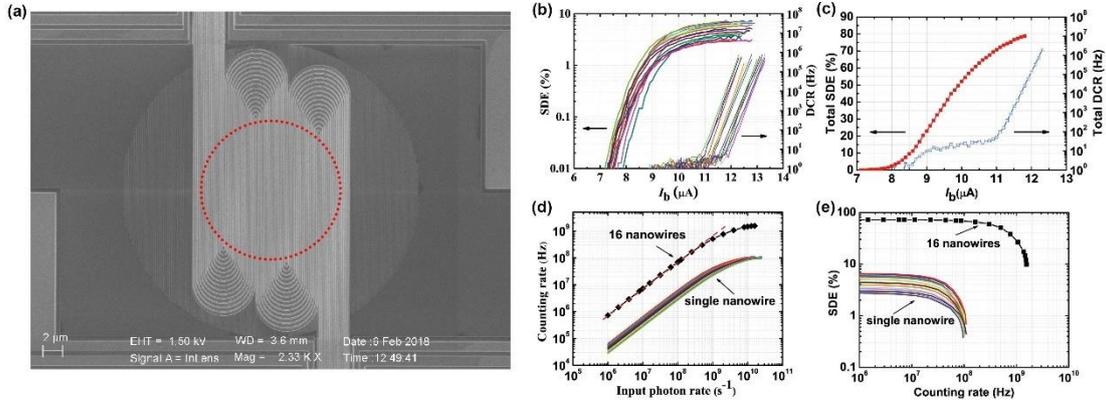

Figure 7. (a) SEM image of a 16-nanowire interleaved SNSPD. (b) SDE and DCR as functions of the bias current for single pixel in the 16-pixel SNSPD. (c) Total SDE and DCR as functions of the bias current of the 16-pixel SNSPD. (d) The measured individual and combined nanowire CRs vs. the input photon rate. The red dashed line is a guide to the eye. (e) The individual and total SDEs as a function of the measured CR. The colored curves represent the results for each nanowire, whereas the dotted curve represents the total SDE obtained by summing for the 16 nanowires. Reproduced with permission. [99] Copyright 2019, IEEE Publishing Ltd.

● Other parameters

Apart from the four aforementioned general parameters, some other parameters are valuable for certain applications.

**Wavelength**: Although an SNSPD is a natural broadband detector due to its low gap energy [103], an optical cavity was integrated to improve the absorption at a certain wavelength, limiting its broadband property. For a specific wavelength from UV (315 and 370 nm) to NIR (up to 2 μm), SNSPDs with high SDE have been reported [80; 104-106]. Broadband SNSPDs with high SDE can be obtained by introducing a broadband cavity or some specific optical structure. Recently interest in the development of broadband and multispectral SNSPDs has been growing [107-109].

**Photon number resolvability (PNR)**: PNR should be good in case of an ideal SPD. However, most SPDs, with the exception of TES, do not have this ability. As a triggering detector, a traditional SNSPD cannot distinguish the photon number. However, with a special readout circuit, SNSPD shows the potential of PNR [110]. Several SNSPDs with quasi-PNR ability were reported on the basis of different space multiplexing methods [99; 111; 112], and a PNR of up to 24 could be observed [113]. If multiple photons were absorbed by a single pixel, then the photon number cannot be resolved.

**Polarization sensitivity**: The classical SNSPD is naturally sensitive to polarization because of its anisotropic meandered nanowire structure. Transverse electric mode photons have a higher absorption than transverse magnetic mode photons. To reduce the polarization sensitivity, various structures, such as spiral, three-dimensional, and fractal structures [79; 114-117], were developed. A polarization extinction ratio (PER) of less than 1.1 was obtained. Some studies adopted the opposite approach to amplify the polarization sensitivity to obtain an SNSPD with a high PER of more than 400 [118-120].

## 3.4 Candidate materials

In the previous two decades, many superconducting materials, including Nb [121], NbN [37], NbTiN [122], NbSi [123], WSi [39], MoSi [40], MoGe [124], MoN [125], TiN [126], and $MgB_2$ [127; 128], have been adopted for fabricating SNSPD. Some attempts used the high-temperature superconductor (HTS) $YBa_2Cu_3O_{7-x}$ [129-131]. The most successful candidates are Nb(Ti)N, WSi, and MoSi, which have an SDE of more than 90%. A general discussion on the materials will be given here instead of a detailed performance comparison, which can be found elsewhere [29].

The selection of suitable candidate materials should ensure that the intrinsic detection efficiency becomes unity. Therefore, the key parameter is $T_c$ (gap energy or Cooper-pair binding energy). Apart from the material itself, other geometric parameters also influence $T_c$, including the thickness of the film and width of the nanowire. A long nanowire structure indicates that SNSPD operates as multiple SNSPDs in series with the same bias current. The performance is bias-current sensitive; hence, nanowire homogeneity is vital with respect to the overall detector performance. For materials with a low $T_c$, a single photon can break more Cooper-pairs; thus, the SNSPD is more sensitive. For the SNSPD made of WSi ($T_c$ = 5 K), obtaining an intrinsic detection efficiency of 1 is easy; this value is registered as a clear saturated plateau in the $SDE$-$I_b$ curve (see Figure 1[a] in [39]). Such a saturated plateau is difficult to obtain in case of the SNSPD comprising NbN ($T_c$ = 16 K) (see Figure 7 in [37]). Strict fabrication requirements (line width and uniformity) may be released partially for SNSPD made using low-$T_c$ materials.

However, the SNSPDs made of low-$T_c$ materials have some disadvantages. Usually, low-Tc materials have a low critical current density (maximum direct current flowing through the superconductors without resistance divided by the cross-section area) and a large kinetic inductance, which cause the SNSPD to have a small output signal amplitude (low SNR→large TJ) and low CR. Another practical issue is the burden on cryogenics. Low $T_c$ indicates high cryocooling costs. Most of the Nb(Ti)N SNSPDs work at temperatures of 2–3 K. However, the SNSPDs made of WSi and MoSi often operate at temperatures lower than 1 K, requiring more complicated and luxurious cryogenics.

Vodolazov recently discussed the detection mechanism of an SNSPD based on the kinetic equation approach [68]. The specific heat capacities of electrons and phonons are the parameters of the candidate materials that can influence the SNSPD detection dynamics. Several micron-wide dirty superconducting bridges can detect a single near-infrared or optical photon. The experimental work on single photon detection with 0.5–5-μm-wide NbN bridge has also been reported [132].

The final question often raised by the users about the materials is the possibility of using HTS such as $YBa_2Cu_3O_{7-x}$. Unfortunately, HTS cannot be used. First, HTS has a large gap energy, resulting in low sensitivity for single-photon detection. Second, as a multi-component compound, achieving good nanoscale homogeneity for a 10-micron-scale or a larger active area is difficult. Third, stability and controllability are challenging to achieve in case of ultrathin HTS films. Registering a single-photon response from HTS nanowires is possible. However, developing a practical SNSPD with an acceptable performance is not realistic using the current material technology.

3.5 Cryogenics

Superconductors have always been associated with cryogenics before room-temperature superconductors are invented. To operate SNSPDs and other superconducting sensors and detectors, the working temperature should be usually lower than 0.5 $T_c$. In case of the Nb(Ti)N SNSPDs, two-stage commercial Gifford–McMahon (G–M) cryocoolers with a cooling power of 0.1 W@4.2 K are widely adopted, which can run continuously at a temperature as low as 2 K with a power consumption of ~1.5 kW [133]. Liquid $^4$He operating at 4.2 K is an alternative, although it is a rare resource with a non-negligible running cost. Reducing the vapor pressure of $^4$He can attain a working temperature lower than 2 K using the superfluid effect. The NbTiN SNSPDs may operate at high temperatures (4–7 K), which is interesting for reducing the cryogenic requirements [104]. Low temperatures are preferable for running WSi and/or MoSi SNSPDs. Sub-1 K cryocoolers, such as the three-stage cryocooler [134], $^3$He refrigerator, adiabatic demagnetization refrigerator, or dilution refrigerator, which are costly and non-portable, were often used.

For scaled application, the size, weight, and power (SWaP) of the cooler cannot be neglected. The SWaP of the two-stage G–M cryocooler is considerably high for a communication base station, let alone the sub-1 K systems. Some studies aimed to reduce the SWaP of a cryocooler for SNSPD using space-application-compatible cryocooler technology [135-138]. Recently, a prototype cryocooler, with a two-stage high-frequency pulse-tube cryocooler and a $^4$He Joule–Thomson cooler driven by linear compressors, reached a minimum temperature of 2.6 K with an input power of 320 W and a weight of 55 kg. The SNSPD hosted by this cryocooler had an SDE of 50% and a TJ of 48 ps [135]. Although this performance is encouraging, there are still several steps to ensure the practicality and cost efficiency of this cooler.

3.6 Niche market

Because SNSPD is a cutting-edge technology, the commercial market is one of the driving forces in improving the performance and making the system practical, user-friendly, and cheap. Six startup companies, namely, ID Quantique (Switzerland) [139], PHOTEC (China) [140], Photon Spot (USA) [141], Quantum Opus (USA) [142], SCONTEL (Russia) [143], and Single Quantum (Netherlands) [144], are working on the commercialization of the SNSPD technology; the first among these companies is SCONTEL, which was founded by Gol'tsman in 2004. The market for this technology is smaller (~20 M USD in 2019) than the market for most semiconducting products. Figure 8 shows the global system installation information. However, the market is expected to continue to grow along with the global investment and commercialization of the QI technologies.

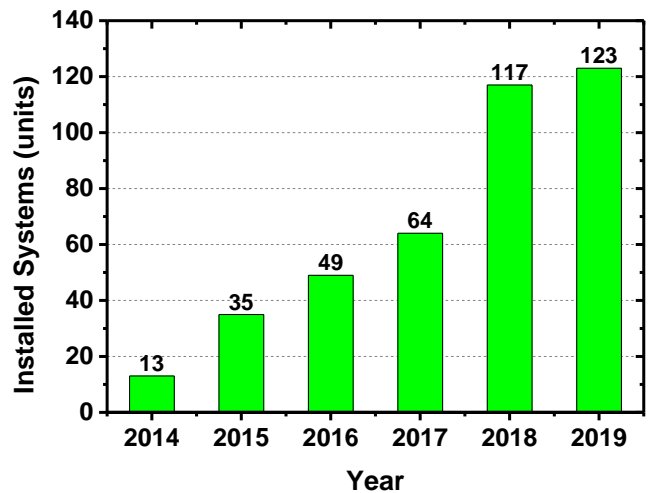

Figure 8 Estimated world market for the sold SNSPD systems (incomplete data from private partner)

4. **Applications in QI**

Since the invention of SNSPD in 2001, it has attracted considerable attention from various research fields. The first application of SNSPD was to diagnose VLSI CMOS circuits in 2013 [145]. The first application in QI started with the characterization of a single-photon source, which was presented by Hadfield et al. [146]. Subsequently, several applications have emerged in different fields such as QI, biological fluorescence [147; 148], deep-space communication [26], and light detection and ranging, including satellite laser ranging [27; 28; 91; 149-153]. The applications in QI are considerably impressive and systematic because QI requires high-performance SPDs. We will introduce the applications in QI in three categories: quantum communication, quantum computation, and others.

4.1 Quantum communication

The InGaAs/InP SPADs dominated in the previously conducted QKD experiments. However, their performance cannot match with the development pace of the QKD because of low DE and high DCR. Until now, the best result obtained using SPADs was reported in 2015, where a QKD containing more than 307 km of optical fiber based on a coherent one-way protocol was demonstrated with the best InGaAs/InP SPADs (DEs of 20%–22%, DCR of ~1 cps) [154]. These detectors were actually cooled to 153 K using a Stirling cryocooler.

The first QKD experiment using SNSPDs was conducted by Hadfield et al. in 2005 [62]. A secure key rate exchange of more than 42.5 km was demonstrated using twin SNSPDs with an SDE of 0.9% and a DCR of 100 Hz. Although the SBER is lower than that when InGaAs SPAD is used, this indicates the considerable potential of using high-performance SNSPD. In 2007, Takesue et al. reported a QKD record of a 12.1-bps secure key rate over 200 km of fiber using SNSPD, which surpassed the distance record achieved using SPAD [155]. Subsequently, more and more QKD experiments have been performed using SNSPDs, effectively improving the QKD distance and key rate. Improvements from single-photon source and other devices and theoretical developments, including new protocols, also contribute to the progress of QKD. Table 2 presents the important experiments of QKD using SNSPD and related parameters (SNSPD parameters, QKD results, and the adopted protocols). Almost all the current transmission distance records of QKD in fiber have been obtained using SNSPDs.

Table 2. List of QKD experiments using SNSPDs

| Year | SNSPD (SDE/DCR) | QKD (Fiber length/SKR) | Protocol | Reference | Note |
|---|---|---|---|---|---|
| 2005 | 0.9%/100 cps | 42.5 km/2 bps | BB84 | [62] | |
| 2007 | 0.7%/<10 cps | 200 km/12.1 bps | DPS | [155] | |
| 2009 | 0.5%/78 cps | 135 km/0.2 bps | Decoy state | [156] | |
| 2010 | 4%/1 cps | 200 km/2.4 bps | Decoy state | [157] | |
| 2011 | 15%/100 cps | 45 km/268.9 kbps | Decoy state | [158] | |
| 2012 | 2.5%/1 cps | 260 km/1.85 bps | DPS | [159] | |
| 2014 | 46%/10 cps | 200 km/0.02 bps | MDI | [160] | |
| 2015 | >40%/10 cps | 30 km/16.9 bps | MDI | [161] | Field test |
| 2015 | 10%/20 cps | 100 km/27.6 bps | Decoy state | [162] | |
| 2015 | 50%/100 cps | 300 km/0.05 bps | MDI | [163] | |
| 2016 | 65%/100 cps | 55 km/16.5 bps | MDI | [164] | Field network |
| 2016 | 65%/30 cps | 404 km/$3.2 \times 10^{-4}$ bps | MDI | [165] | Ultralow-loss fiber |
| 2017 | N/A | 80 km/100 bps | MDI | [166] | |
| 2018 | 40%–60%/0.1 cps | 421.1 km/0.25 bps | Decoy state | [167] | Ultralow-loss fiber |
| 2019 | 46%/10 cps | 100 km/14.5 bps | MDI | [168] | Asymmetric channels |
| 2019 | 53%/50 cps | 140 km/268 bps 180 km/31 bps | MDI | [169] | |
| 2019 | 75%/100 cps | 300 km/39.2 bps | TF | [170] | |
| 2019 | 60.8%/95 cps | 300 km/2.01 kbps | TF | [171] | Asymptotic key rate |
| 2019 | 44%/22 cps | 454 km/0.045 bps | TF | [172] | Asymptotic key rate |
| 2020 | 40%/10 cps | 502 km/0.118 bps | TF | [64] | Ultralow-loss fiber |
| 2020 | 57%/3.5 cps | 509 km/0.269 bps | TF | [65] | Ultralow-loss fiber |

* DPS: Differential phase shift

### 4.2 Quantum computation

Before 2017, majority of the optical quantum computation experiments were performed at wavelengths of approximately 800 or 900 nm, where high-quality single-photon sources and SPDs (SPAD: DE of ~60% for 800 nm and ~30% for 900 nm) are available [173-176]. Further developments acquired SPDs with high SDE and repetition rate, which cannot be achieved using SPADs. In 2017, He et al. demonstrated four-photon boson sampling using two SNSPDs (SDE of 52% at 910-nm wavelength and CR of 12.9 MHz), which surpassed the previous results over 100 times [177]. Then, Wang et al. demonstrated scalable boson sampling with photon loss, where 13 SNSPDs were adopted [178]. Furthermore, using 24 SNSPDs with an SDE of 75% at a wavelength of 1550 nm, Zhong et al. presented first 12-photon genuine entanglement with a state fidelity of $0.572 \pm 0.024$ [179].

The same group recently developed solid-state sources containing highly efficient, pure, and indistinguishable single photons and 3D integration of ultralow-loss optical circuits. The team performed experiments by feeding 20 pure single photons into a 60-mode interferometer (60 SNSPDs with an SDE of 60%–82%) (see Figure 2 for the schematics). This result yielded an output with a $3.7 \times 10^{14}$-dimensional Hilbert space, more than 10 orders of magnitude larger than those obtained in previous experiments, which, for the first time, enters into a genuine sampling regime, where it becomes impossible to exhaust all the possible output combinations [60]. The results were validated against distinguishable and uniform samplers with a confidence level of 99.9%. This result was equivalent to 48 qubits and approached a milestone for boson sampling [180]. We believe that quantum supremacy/advantage with boson sampling will be achieved soon using high-performance SNSPDs.

### 4.3 Other QI applications.

In the previous 15 years, apart from QKD and optical quantum communication, numerous advanced QI applications that use SNSPD have been demonstrated, such as quantum teleportation [181; 182], quantum storage [183; 184], quantum money, quantum switch, quantum digital signature, quantum fingerprint, quantum data lock, quantum ghost imaging, quantum time transfer, and quantum entropy. For lack of space, not all the achievements can be presented in this review. Therefore, some selected results are introduced here.

**Bell's inequalities validation:** In 1964, John Stewart Bell, discovered inequalities that allow an experimental test of the predictions of local realism against those of standard quantum physics [185], which can answer the question about the completeness of the formalism of quantum mechanics raised by Albert Einstein, Boris Podolsky, and Nathan Rosen (also known as the EPR paradox) [186]. In the ensuing decades, experimentalists performed increasingly sophisticated tests with respect to Bell's inequalities. However, these tests have always contained at least one "loophole," allowing a local realist interpretation of the experimental results [187]. In 2015, by simultaneously closing two main loopholes, three teams independently confirmed that we must definitely renounce local realism [63; 188; 189]; one team at National Institute of Standards and Technologies (NIST) closed the detection loophole using SNSPDs [63].

The NIST experiment was based on the scheme presented in Figure 9 [187]. The team used rapidly switchable polarizers (A and B) located more than 100 m from the source to close the locality loophole. WSi SNSPDs with an SDE of 91% ± 2% were required to close the detection loophole. Pairs of photons were prepared using a nonlinear crystal to convert a pump photon into two "daughter" entangled photons ($v_1$ and $v_2$). Each photon was sent to a detection station with a polarizer, the alignment of which was randomly established. The team achieved an unprecedented high probability that when a photon enters one analyzer, its partner enters the opposite analyzer. Based on this and the high-efficiency SNSPDs, a heralding efficiency of 72.5% can be achieved, which was larger than the critical value of 2/3. The confidence level of the measured violation of Bell's inequality can be evaluated by the probability $p$ that a statistical fluctuation in a local realist model would yield the observed violation. The reported $p$ is $2.3 \times 10^{-7}$, which corresponded to a violation by 7 standard deviations. The results firmly establish several fundamental QI schemes such as device-independent quantum cryptography and quantum networks.

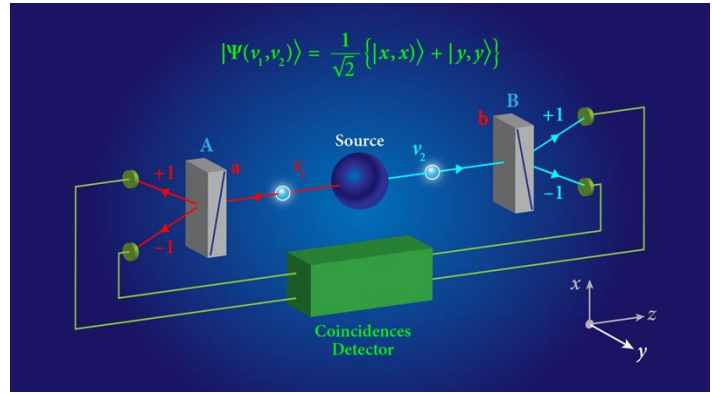

Figure 9 The apparatus for performing the Bell test. A source emits a pair of entangled photons $v_1$ and $v_2$. Their polarizations are analyzed by polarizers A and B (grayblocks) aligned along the directions a and b. Each polarizer has two output channels labeled as +1 and −1. The final correlation of the photons can be determined by the coincidence detectors (SNSPDs). (APS/Alan Stonebraker)

**Quantum random number generation:** Randomness is important for many information processing applications. Device-independent quantum random number generation based on the loophole-free violation of a Bell inequality is the objective in QI science. Liu et al. used the state-of-the-art quantum optical technology to create, modulate, and detect the entangled photon pairs, achieving an efficiency of more than 78% from creation to detection at a distance of approximately 200 m; two SNSPDs with SDEs of more than 92% were obtained at a wavelength of 1550 nm [190]. Then, $6.25 \times 10^7$ quantum-certified random bits were obtained in 96 h with a total failure probability of less than $10^{-5}$. This achievement is a crucial step toward a key aspect of practical applications that require high security and genuine randomness.

**Integrated quantum photonics (IQP)**: IQP is an important field in quantum information, which may provide an integrated platform for almost all the optical QI applications. A single-photon detector is an indispensable technology for IQP. Pernice et al. first demonstrated the integration of SNSPD on a silicon waveguide, which had an on-chip DE of 91% [73]. Subsequently, the SNSPDs on waveguides made of various materials, such as SiN [191; 192], GaAs [193; 194], AlN [195], LiNbO$_3$ [196], and diamond [197], were demonstrated, paving the way for all types of IQP applications that use SNSPDs. We refer the readers to a recent review [32] and references therein for details.

## 5. Summary and outlook

Because of the extensive development of QI, SNSPDs have progressed rapidly in terms of science and technology as well as application. A niche market is available and growing continuously. On the contrary, high-performance SNSPDs have been advancing many fields, including but not limited to QI. However, there is considerable room to further improve the SNSPDs. Some of the related points are listed below.

1. Performance improvement: Users always want to have SNSPDs with better performances. The requirements for all-round SNSPDs are increasing, which require that two or more parameter requirements should be simultaneously satisfied. For example, TF–QKD requires SNSPDs with high SDE and low DCR. Boson sampling needs many SNSPDs with high SDE and CR. Special applications need unique SNSPDs for other wavelengths, such as mid-infrared or longer wavelengths, broadband SNSPD, and SNSPD with the PNR ability.
2. Array technology: Fabricating a single-pixel SNSPD is easier than fabricating an SNSPD array. An array acquires films with high homogeneity and a nanofabrication process with a high uniformity. The readout in case of an SNSPD array is another key technology for array application because the detectors are operated at considerably low temperatures. Some inspiring results and studies have been recently observed [98; 198]. For example, the first kilopixel SNSPDs with row-column multiplexing architecture were demonstrated in 2019 [98]. However, their performance, such as efficiency and uniformity, should be further improved.
3. Cryogenics: SWaP and the price of the system decide whether this technology can be massively applied. SWaP and the price of the detectors can be considerably reduced by improving their yield and performance. To a certain degree, cryogenics will determine the future of SNSPDs. We need to design and develop more compact, portable, and affordable cryocoolers customized for SNSPDs. Finally, the extent of this development will be dependent on the commercialization and industrialization of QI.

Acknowledgment

The author would like to thank Mengting Si for reading the manuscript and providing fruitful comments and suggestions. This work was supported by the National Key R&D Program of China under grant no. 2017YFA0304000, the National Natural Science Foundation of China under grant no. 61671438, Shanghai Municipal Science and Technology Major Project under grant no. 2019SHZDZX01, and the Program of Shanghai Academic/Technology Research Leader under grant no. 18XD1404600. This manuscript is also dedicated to the doctors and nurses fighting the coronavirus in China when the manuscript was prepared.